\documentclass[elsart12,epsf]{revtex4}
\begin{document}

% ABBREVIATED JOURNAL NAMES  

%

\def\ap#1#2#3{     {\it Ann. Phys. (NY) }{\bf #1} (19#2) #3}

\def\arnps#1#2#3{  {\it Ann. Rev. Nucl. Part. Sci. }{\bf #1} (19#2) #3}

\def\npb#1#2#3{    {\it Nucl. Phys. }{\bf B#1} (19#2) #3}

\def\plb#1#2#3{    {\it Phys. Lett. }{\bf B#1} (19#2) #3}

\def\prd#1#2#3{    {\it Phys. Rev. }{\bf D#1} (19#2) #3}

\def\prep#1#2#3{   {\it Phys. Rep. }{\bf #1} (19#2) #3}

\def\prl#1#2#3{    {\it Phys. Rev. Lett. }{\bf #1} (19#2) #3}

\def\ptp#1#2#3{    {\it Prog. Theor. Phys. }{\bf #1} (19#2) #3}

\def\rmp#1#2#3{    {\it Rev. Mod. Phys. }{\bf #1} (19#2) #3}

\def\zpc#1#2#3{    {\it Z. Phys. }{\bf C#1} (19#2) #3}

\def\mpla#1#2#3{   {\it Mod. Phys. Lett. }{\bf A#1} (19#2) #3}

\def\nc#1#2#3{     {\it Nuovo Cim. }{\bf #1} (19#2) #3}

\def\yf#1#2#3{     {\it Yad. Fiz. }{\bf #1} (19#2) #3}

\def\sjnp#1#2#3{   {\it Sov. J. Nucl. Phys. }{\bf #1} (19#2) #3}

\def\jetp#1#2#3{   {\it Sov. Phys. }{JETP }{\bf #1} (19#2) #3}

\def\jetpl#1#2#3{  {\it JETP Lett. }{\bf #1} (19#2) #3}

%%%%%%%%% notice the parenthesys is only on one side

\def\ppsjnp#1#2#3{ {\it (Sov. J. Nucl. Phys. }{\bf #1} (19#2) #3}

\def\ppjetp#1#2#3{ {\it (Sov. Phys. JETP }{\bf #1} (19#2) #3}

\def\ppjetpl#1#2#3{{\it (JETP Lett. }{\bf #1} (19#2) #3} 

\def\zetf#1#2#3{   {\it Zh. ETF }{\bf #1}(19#2) #3}

\def\cmp#1#2#3{    {\it Comm. Math. Phys. }{\bf #1} (19#2) #3}

\def\cpc#1#2#3{    {\it Comp. Phys. Commun. }{\bf #1} (19#2) #3}

\def\dis#1#2{      {\it Dissertation, }{\sf #1 } 19#2}

\def\dip#1#2#3{    {\it Diplomarbeit, }{\sf #1 #2} 19#3 }

\def\ib#1#2#3{     {\it ibid. }{\bf #1} (19#2) #3}

\def\jpg#1#2#3{        {\it J. Phys}. {\bf G#1}#2#3}  

%

%\hoffset-1cm

% Yale printer values
\voffset1.5cm

%\draft
%\preprint{hep-ph/}

\title{From target to projectile and back again: selfduality of high energy evolution.}
\author{Alex Kovner and  Michael Lublinsky}

\address{Physics Department, University of Connecticut, 2152 Hillside
Road, Storrs, CT 06269-3046, USA}
\date{\today}

\begin{abstract}
We prove that the complete kernel for the high energy evolution in QCD must be selfdual. The relevant duality transformation is formulated in precise mathematical terms and is shown to transform the charge density into the functional derivative with respect to the single-gluon scattering matrix. This transformation interchanges the high and the low density regimes.
We demostrate that the original JIMWLK kernel, valid at large density is indeed dual to the low denisity limit of the complete kernel derived recently in hep-ph/0501198.

\end{abstract}
\maketitle
\setcounter{equation}{0}

%%%%%%%%%%%%%%%%%%%%%%%%%%%%%%%%%%%%%%%%%%%%%%%%%%%%%%%%%%
Recent months have seen renewed attempts to understand Pomeron loops contribution to the high energy evolution of hadronic cross section in QCD \cite{pomeron}. Over the years much work has been done to understand the evolution in the reggeon language\cite{reggeon}. 
In recent years the study of the high energy scattering has centered around the so called JIMWLK evolution equation \cite{balitsky,JIMWLK,cgc} and its mean field limit \cite{kovchegov}. These describe the approach of the scattering amplitude to saturation due to multiple scattering corrections on dense hadronic targets, or in the diagrammatic language, the fan diagrams.
The JIMWLK equation however only partially takes into account the processes whereby the gluons emitted in the projectile wave function at an early stage of the evolution, are "bleached" by subsequently emitted gluons, or the so-called Pomeron loops.

 The work \cite{im1,ms} lead to realization of the absence 
of the Pomeron loops in the JIMWLK evolution equation
and the resulting violation of $t$-channel unitarity. 
Very recently several papers have derived corrections to the JIMWLK evolution equation away from the high density limit \cite{ploops,kl}. The most complete of the results so far is given in \cite{kl} where the kernel for the evolution equation in the low density limit is derived. This kernel has a very specific form which suggests an intriguing duality between the high density and low density limit.

In the present paper we prove that such a duality of the evolution kernel is in fact required by the $t$-channel unitarity and the Lorentz invariance of the theory. We consider the evolution from the projectile side and the target side separately.
The requirement that the evolution of the projectile and the target wave functions has the same functional form \cite{footnote}  coupled with the requirement of Lorentz invariance of the scattering matrix, leads to the condition that the kernel of the evolution $\chi[\rho,{\delta\over\delta\rho}]$ must be invariant under the transformation
\begin{equation}
\rho\rightarrow -i{\delta\over\delta\alpha}; \ \ \ \ \ \ \ {\delta\over\delta\rho}\rightarrow i\alpha
\label{duality}
\end{equation}
where $\rho$ is the charge density in the target wave function, and $\alpha$ is the phase of the  
scattering matrix of a single gluon on the same target (see below). The discussion of this paper assumes the eikonal form of the $S$-matrix, and is thus valid in the leading logarithmic approximation.

We start by considering a general expression for the $S$-matrix of a projectile with the wave function $|P\rangle$ on a target with the wave function $|T\rangle$\cite{kl1}, where the total rapidity of the process is $Y$. The projectile is assumed to be moving to the left with total rapidity $Y-Y_0$ (and thus has sizeable color charge density $\rho^-$), while the target is moving to the right
 with total rapidity $Y_0$ (and has large $\rho^+$). We assume that the projectile and the target 
contain only partons with large $k^-$ and $k^+$ momenta respectively: $k^->\Lambda^-$ and $k^+>\Lambda^+$. 
The eikonal expression for the $S$-matrix reads
\begin{equation}
{\cal S}_{Y}\,\,=\,\,\int\, D\rho^{+a}(x,x^-)\,\, W^T_{Y_0}[\rho^+(x^-,x)]\,\,\Sigma^P_{Y-Y_0}[\alpha]\,,
\label{s}
\end{equation}
where $\Sigma^P$ is the $S$-matrix averaged over the projectile wave function\cite{im}
\begin{equation}
\Sigma^P[\alpha]\,\,=\,\,\langle P|\,{\cal P} e^{i\int dx^-\int d^2x\hat\rho^{-a}(x)\alpha^a(x,x^-)}\,|P\rangle\,.
\label{sigma}
\end{equation}
($\cal P$ denotes path ordering with respect to $x^-$),
and $W^T[\alpha]$ is the weight function representing the target, 
which is related to the target wave function in the following way: for an arbitrary operator $\hat O[\hat\rho^+]$
\begin{equation}
\langle T|\,\hat O[\hat\rho^+(x)]\,|T\rangle\,\,=\,\,\int\, D\rho^{+a}\,\, W^T[\rho^+(x^-,x)]\,\,O[\rho^+(x,x^-)]\,.
\label{w}
\end{equation}
The field $\alpha(x)$ is the $A^+$ component of the vector potential in the light cone gauge $A^-=0$. This is the natural gauge from the point of view of partonic interpretation of the projectile wave function. In the same gauge the target charge density $\rho^+$ is related to $\alpha$ through the solution of the classical equations of motion\cite{JIMWLK,cgc}
\begin{equation}
\alpha^a(x,x^-)T^a\,\,=\,\,{1\over \partial^2}(x-y)\,
\left\{U^\dagger(y,x^-)\,\,\rho^{+a}(y,x^-)\,T^a\,\,U(y,x^-)\right\}, \ \ \ \ \ 
U(x,x^-)\,\,=\,\,{\cal P}\,\exp\{i\int_{-\infty}^{x^-} dy^-T^a\alpha^a(x,y^-)\}\,.
\end{equation}
where $T^a_{bc}=if^{abc}$ is the generator of the $SU(N)$ group in the adjoint representation.
The unitary matrix $U(x,x^-\rightarrow\infty)$ has the meaning of the scattering matrix of a single gluon from the projectile wave function on the target $|T\rangle$.

In the formulae above we use hats to denote quantum operators. Note that the quantum operators 
\begin{equation}
\hat\rho^{-a}(x)=\int_0^{\infty} dk^-a^{\dagger b}_i(x,k^-) T^a_{bc}a^c_i(x,k^-),\ \ \ \ \ \hat\rho^{+a}(x)=\int_0^{\infty} dk^+\bar a^{\dagger b}_i(x,k^+) T^a_{bc}\bar a^c_i(x,k^+)
\end{equation}
 do not depend on longitudinal coordinates, but only on transverse coordinates $x$ (here $a^\dagger$ and $\bar a^\dagger$ stand for single gluon creation operators of left and right moving gluons respectively). 
The "classical" variables $\alpha$ and $\rho^+$ on the other hand do depend on the longitudinal coordinate $x^-$. This dependence, as discussed in detail in \cite{kl} arises due to the need to take correctly into account the proper ordering of noncommuting quantum operators. Thus the ordering of the quantum operators $\hat\rho^+$ in the expansion of $\hat O$ in the lhs of eq.(\ref{w}) translates into the same ordering with respect to the longitudinal coordinate $x^-$ of $\rho^+(x^-)$ in the expansion of $O[\rho^+(x^-)]$ in the rhs of eq.(\ref{w}). 

As was shown in \cite{kl} the functional $W^T[\alpha]$ cannot in general be interpreted as probability density, as it contains a complex factor. This factor - the Wess-Zumino term, ensures correct commutators between the quantum operators $\hat\rho^a$. In the present paper we do not require an explicit form of this term, but the following property which is implicit in eq. (\ref{w}) is crucial to our discussion. The "correlators" of the charge density $\langle \rho^{a_1}(x_1,x^-_1)... \rho^{a_n}(x_n,x^-_n)\rangle$ do not depend on the values of the longitudinal coordinates $x^-_i$, but only on their ordering\cite{kl}.

Note that one can define an analog of $W^T$ for the wave function of the projectile via
\begin{equation}
\langle P|\,\hat O[\hat\rho^-(x)]\,|P\rangle\,\,=\,\,\int \,D\rho^{-a}\,\, W^P[\rho^-]\,\,O[\rho^-(x,x^-)]\,.
\label{wp}
\end{equation}
With this definition it is straightforward to see that $\Sigma^P$ and $W^P$ are related through a 
functional Fourier transform.
To represent $\Sigma$ as a functional integral with weight $W^P$ we have to order the factors of the charge density $\hat\rho^-$ in the expansion of  eq.(\ref{sigma}), and then endow the charge density $\hat\rho^-(x)$ with an additional coordinate $t$ to turn it into a classical variable. This task is made easy by the fact that the ordering of $\hat\rho$ in eq.(\ref{sigma}) follows automatically the ordering of the coordinate $x^-$ in the path ordered exponential. Since the correlators of $\rho(x,t_i)$ with the weight $W^P$ depend only on the ordering of the coordinates $t_i$ and not their values, we can simply set $t=x^-$. 
Once we have turned the quantum operators $\hat\rho$ into the classical variables $\rho(x^-)$, the path ordering plays no role anymore, and we thus have
\begin{equation}
\Sigma^P(\alpha)\,\,=\,\,\int\, D\rho^{a}\,\, W^P[\rho]\,\,e^{i\int dx^-\int d^2x\rho^{a}(x,x^-)\alpha^a(x,x^-)}.
\label{sw}
\end{equation}

We now turn to the discussion of the evolution.
The evolution to higher energy can be achieved by  boosting either the projectile or the target. The resulting $S$-matrix should be the same. This is required by the Lorentz invariance of the $S$-matrix.
Consider first boosting the projectile by a small rapidity $\delta Y$. 
This transformation leads to the change of the projectile $S$-matrix $\Sigma$ of the form
\begin{equation}
{\partial\over\partial Y}\,\Sigma^P\,\,=\,\,\chi^\dagger[\alpha,{\delta\over\delta\alpha}]\,\,\Sigma^P[\alpha]
\label{dsigma}
\end{equation} 
The explicit form of the kernel $\chi$ is known in the large density limit \cite{JIMWLK},\cite{cgc} and 
in the small density limit \cite{kl}.
Substituting eq.(\ref{dsigma}) into eq.(\ref{s}) we have
\begin{eqnarray}
{\partial\over\partial Y}\,{\cal S}_{Y}&=&\int \,D\rho^{+a}(x,x^-)\,\, W^T_{Y_0}[\rho^+(x^-,x)]\,\,
\left\{\chi^\dagger[\alpha,{\delta\over\delta\alpha}]\,\,\Sigma^P_{Y-Y_0}[\alpha]\right\}\nonumber\\
&=&
\int \,D\rho^{+a}(x,x^-)\,\,\left \{\chi[\alpha,{\delta\over\delta\alpha}]\,\,
W^T_{Y_0}[\rho^+(x^-,x)]\right\} \,\,\Sigma^P_{Y-Y_0}[\alpha]\,.
\label{ds}
\end{eqnarray}
Where the second equality follows by integration by parts. 
We now impose the requirement that the $S$-matrix does not depend on $Y_0$
\cite{LL}. Since $\Sigma$ 
in eq.(\ref{s}) depends on the difference of rapidities, 
eq.(\ref{dsigma}) implies
\begin{equation}
{\partial\over\partial Y_0}\,\Sigma\,\,=\,\,-\,
\chi^\dagger[\alpha,{\delta\over\delta\alpha}]\,\,\Sigma[\alpha]\,.
\label{dsigma0}
\end{equation} 
Requiring that $\partial{\cal S}/\partial Y_0\,=\,0$ 
we find that $W$ should satisfy 
\begin{equation}
{\partial\over\partial Y}\,W^T\,\,=\,\,
\chi[\alpha,{\delta\over\delta\alpha}]\,\,W^T[\rho^+]
\label{dwt}
\end{equation}

Thus we have determined the evolution of the target eq.(\ref{dwt}) by boosting the projectile
and requiring Lorentz invariance of the $S$-matrix.
On the other hand the extra energy due to boost can be deposited in the target rather than in the projectile. 
How does $W^T$ change under boost of the target wave function? To answer this question
we consider the relation between $\Sigma$ and $W$  together with the evolution of $\Sigma$.
Referring to eqs.(\ref{sw}) and (\ref{dsigma}) it is obvious that multiplication of $\Sigma^P$ by $\alpha$ is 
equivalent to acting on $W^P$ by the operator $-i\delta/\delta\rho$, and acting on $\Sigma^P$ by 
$\delta/\delta\alpha$ is equivalent to multiplying $W^P$ by $i\rho$. Additionally, the action of 
$i\rho$ and $-i\delta/\delta\rho$ on $W^P$ must be in the reverse order to the action of 
$\delta/\delta\alpha$ and  $\alpha$ on $\Sigma^P$. This means that the evolution of the functional 
$W^P$ is given by
\begin{equation}
{\partial\over\partial Y}\,W^P\,\,=\,\,\chi[-i{\delta\over\delta\rho},i\rho]\,\,W^P[\rho]\,.
\label{dwp}
\end{equation}
Although eq.(\ref{dwp}) refers to the weight functional representing the projectile wave function, the $t$-channel unitarity requires that the functional form of the evolution must be the same for $W^T$. Comparing eq.(\ref{dwt}) and eq.(\ref{dwp}) we find that the high energy evolution kernel must, as advertised,  satisfy the selfduality relation
\begin{equation}\label{main}
\chi[\alpha,\,{\delta\over\delta\alpha}]\,\,=\,\,\chi[-i{\delta\over\delta\rho},\,i\rho]\,.
\end{equation}
This is the main result of the present paper.

It is interesting that although we do not know the complete kernel $\chi$, we can check that the known limits of it are consistent with the duality eq.(\ref{duality}).
In particular, the duality transformation interchanges the large and small density limits, and the kernel $\chi$ is known in both these limits. In the large density limit the pertinent expression is the JIMWLK kernel. It can be conveniently written as \cite{kl1}
\begin{eqnarray}\label{JIMWLK}
\chi_{\rho\rightarrow\infty}&=&
\frac{\alpha_s}{2\pi^2}\int_{x,y,z} {(z-x)_i(z-y)_i
\over (z-x)^2(z-y)^2}\left\{ {\delta\over \delta \alpha^a(x,-\infty)}{\delta\over \delta \alpha^a(y,-\infty)}+ {\delta\over \delta \alpha^a(x,\infty)}{\delta\over \delta \alpha^a(y,\infty)}\right.\\
&-& 2\,\left.{\delta\over \delta \alpha^a(x,-\infty)}\left[{\cal P}e^{i\int_{-\infty}^{\infty} d x^-T^c\alpha^c(x^-,z)}\right]^{ab}{\delta\over \delta \alpha^b(y,\infty)}\right\}\,.\nonumber
\end{eqnarray}
The low density limit of the kernel has recently been calculated in \cite{kl}. It reads
\begin{eqnarray}\label{KL}
\chi_{\rho\rightarrow 0}=&-&
\frac{\alpha_s}{2\pi^2}\int_{x,y,z} {(z-x)_i(z-y)_i
\over (z-x)^2(z-y)^2}\,\{\rho^a(x,-\infty)\rho^a(y,-\infty)+  \rho^a(x,\infty)\rho^a(y,\infty)\\
&-&2\,  \rho^a(x,-\infty)\left[{\cal P}e^{\int_{-\infty}^{\infty} d x^-T^c{\delta\over\delta\rho^c(x^-,z)}}\right]^{ab}\rho^b(y,\infty)\}\,.\nonumber
\end{eqnarray}
Obviously the kernel eq.(\ref{JIMWLK}) transforms into the kernel eq.(\ref{KL}) by the duality transformation eq.(\ref{duality}).

One can go one step further and consider also preasymptotic terms. In \cite{kl1} we have derived a set of preasymtotic corrections to eq.(\ref{JIMWLK}) (not a complete set) which take into account higher density terms in the projectile wave function:
\begin{eqnarray}
\chi_{{\rm large}\ \rho}&=&
\frac{1}{2\pi}\int_z\left\{ b^a_i(z,x^-\rightarrow-\infty,[{\delta\over \delta \alpha}])b^a_i(z,x^-\rightarrow-\infty,[{\delta\over \delta \alpha}])+ 
 b^a_i(z,x^-\rightarrow\infty,[{\delta\over \delta \alpha}])b^a_i(z,x^-\rightarrow\infty,[{\delta\over \delta \alpha}])
\right.\nonumber\\
&-&2\, \left. b^a_i(z,x^-\rightarrow-\infty,[{\delta\over \delta \alpha}])\left[{\cal P}e^{i\int_{-\infty}^{\infty} d x^-T^c\alpha^c(x^-,z)}\right]^{ab}b^b_i(z,x^-\rightarrow\infty,[{\delta\over \delta \alpha}])\right\}
\label{notlarge}
\end{eqnarray}
The field $b^a_i$ satisfies the "classical" equation of motion
\begin{eqnarray}\label{b}
&&\left\{\partial_i+gf^{abc}b^c_i(z,[\rho])\right\}\partial^+ b_i^b(z,[\rho])=g\rho^a(z,x^-)\,;\nonumber\\
&&\epsilon_{ij}[\partial_ib^a_j(z,[\rho])-\partial_jb^a_i(z,[\rho])+gf^{abc}b^b_i(z,[\rho])b^c_j(z,[\rho])]=0\,.
\end{eqnarray}
A correction to the low density limit was derived in \cite{kl}, resumming the same type of corrections but in the target wave function. The relevant expression is
\begin{eqnarray}\label{notsmall}
\chi_{{\rm small}\ \rho}=&-&
\frac{1}{2\pi}\int_{z} \{ b^a_i(z,x^-\rightarrow-\infty,[\rho])b^a_i(z,x^-\rightarrow-\infty,[\rho])
+b^a(z,x^-\rightarrow\infty,[\rho])b^a(z,x^-\rightarrow\infty,[\rho])
\nonumber \\
&-& 2\,b^a_i(z,x^-\rightarrow-\infty,[\rho])\left[{\cal P}e^{\int_{-\infty}^{\infty} d x^-T^c{\delta\over\delta\rho^c(x^-,z)}}\right]^{ab}b^b_i(z,x^-\rightarrow\infty,[\rho])\}\,.
\end{eqnarray}
Again, we see that the two expressions eqs.(\ref{notlarge}) and (\ref{notsmall}) are related by the duality transformation eq.(\ref{duality}). 

In this Letter we have established the exact selfduality property that has to be satisfied by the complete kernel of the high energy evolution in the leading logarithmic approximation. The existing expressions of the limits of $\chi$  are transformed into each other by the duality transformation, as they should.  
We hope that this property will be helpful in future work aimed at finding the complete kernel. The selfduality must be preserved by any expression which purports to contain all Pomeron loops.
In fact, one can suggest an expression based on the known large and small density asymptotics:
\begin{eqnarray}\label{selfdual}
\chi&=&
\frac{\alpha_s}{2\pi^2}\int_{x,y,z} {(z-x)_i(z-y)_i
\over (z-x)^2(z-y)^2}\left\{\left[ {\delta\over \delta \alpha^a(x,-\infty)}+i\rho^a(x,-\infty)\right]\left[{\delta\over \delta \alpha^a(y,-\infty)}+i\rho^a(y,-\infty)\right]\right.\\
&+& \left[{\delta\over \delta \alpha^a(x,\infty)}+i\rho^a(x,\infty)\right]\left[{\delta\over \delta \alpha^a(y,\infty)}+i\rho^a(y,\infty)\right]\nonumber\\
&-& 2\,\left.\left[{\delta\over \delta \alpha^a(x,-\infty)}+i\rho^a(x,\infty)-\right]\left[Pe^{i\int_{-\infty}^{\infty} d x^-T^c\left\{\alpha^c(x^-,z)-i{\delta\over\delta\rho^c(x^-,z)}\right\}}\right]^{ab}\left[{\delta\over \delta \alpha^b(y,\infty)}+i\rho^b(y,\infty)\right]\right\}\,.\nonumber
\end{eqnarray}
This provides an interpolation between the large and small field limits, and preserves selfduality. Of course, this interpolation is not unique, and the exact kernel may turn out to be different. The expression eq.(\ref{selfdual}) has however a simple property, that it becomes the original JIMWLK expression rotated into a different basis. Physically the meaning of the interpolation eq.(\ref{selfdual}) can be understood by 
recalling that $i\delta/\delta\alpha$ has the meaning of $\rho^-$ \cite{kl1}. Thus eq.(\ref{selfdual}) suggests that the physics does not depend on the current densities $\rho^+$ and $\rho^-$ separately, but only on the total color charge density in the lab frame $\rho^0=\rho^-+\rho^+$.

The selfduality of the kernel is somewhat similar (although different in detail) to the duality symmetry of a harmonic oscillator Hamiltonian $p\rightarrow x,\ \ x\rightarrow -p$. One thus hopes that it may eventually be of help in solving the complete evolution equation, once it is derived.

Finally we would like to mention that Lipatov`s 
effective action approach \cite{Lipatov} to the high energy evolution
in QCD is claimed to respect Lorentz invariance and both 
the direct and cross channel unitarity. It would be interesting to
 relate Lipatov's effective action as well as reggeon field theory
directly to our present approach.


\begin{thebibliography}{99}
\bibitem{footnote} The requirement that the evolution of the target and the projectile wave functions be identical is directly related to the $t$-channel unitarity, as it leads to the appearance of the same amplitude above and below a $t$-channel cut. It is only in this sense that the result of this paper relies on $t$-channel unitarity.
\bibitem{im} Note that we are not assuming here that the fields of the projectile or target are small. The weak field limit of this formula has been used in the framework of the dipole model in Ref. \cite{im1}.
\end{thebibliography}
\end{document}